\documentclass[doublecol]{./epl2} 
\usepackage{amssymb}
\usepackage{amsmath}
\newcommand{\beq}{\begin{equation}}
\newcommand{\enq}{\end{equation}}
\newcommand{\bea}{\begin{eqnarray}}
\newcommand{\ena}{\end{eqnarray}}
\newcommand{\rr}{\mathbf r}
\newcommand{\ad}{a^{\dag}}
\newcommand{\butterflycitation}{\cite{niemeyer1999,jaksch2003,oktel2007,goldbaum2008}}
\begin{document}

\title{Rotating states for trapped bosons in an optical lattice}
\author{Emil Lundh}
\institute{Department of Physics, Ume{\aa} University, 
SE-90187 Ume{\aa}, Sweden}

\abstract{
Rotational states for trapped bosons in an optical lattice 
are studied in the framework of the Hubbard model. Critical 
frequencies are calculated and the main parameter regimes 
are identified. Transitions are observed from edge superfluids  
to vortex lattices with Mott insulating cores, and 
subsequently to lattices of interstitial vortices. The former 
transition coincides with the Mott transition. 
Changes in symmetry of the vortex lattices are observed as a function 
of lattice depth.
Predictions for experimental signatures are presented.
}
\maketitle

Ultracold atoms in optical lattices combined with trapping potentials 
offer rich possibilities for exploring new physics. In particular, 
the presence of a trap 
allows for the coexistence of phases 
such as the Mott insulator and superfluid, as was predicted by 
Jaksch {\it et al.} \cite{jaksch1998}, realized by Greiner {\it et al.} 
\cite{greiner2002}, and directly verified 
by two independent groups in 2006 \cite{campbell2006,folling2006}.
The harmonic confinement results in a shell-like structure, where 
superfluid and Mott insulating regions occupy alternating 
intervals (in one dimension), annuli (in two dimensions), or 
spherical shells (in three dimensions). 

Motivated by the progress in the study of vortex physics 
in trapped condensates, rotational 
states in optical lattices have attracted some recent attention. 
At one end of the rich spectrum of possibilities is the case of a 
weak periodic potential that acts as a lattice of pinning centers for 
vortices; this has been realized in experiment using a corotating 
periodic potential \cite{reijnders2004,tung2006}. For deeper optical 
lattices, one may consider a discretized description of the gas, 
the bosonic Hubbard model \cite{jaksch1998,bhat2006b}, and study 
how rotational 
states interact with the quantum phases of the model. It has been 
predicted that quantized vortices in the Hubbard model possess a 
Mott insulating core \cite{wu2004,goldbaum2008}, and also that 
in the limit of a deep lattice, 
rotational states appear as an edge superfluid surrounding a Mott 
insulator \cite{scarola2007}. In the limit of rapid rotation, when the number 
of vortices is comparable to the number of lattice sites, intriguing 
phenomena are predicted such as strongly correlated vortex liquids 
\cite{burkov2006} and fractal densities of states, 
the so-called 
Hofstadter butterfly \butterflycitation.

This study aims to investigate the interplay 
between the quantum phase transition, rotation, and trapping for atoms in 
a optical lattice described by the bosonic Hubbard model, 
and to describe the different types of 
rotational state that can be found in such a system. It will be 
seen that at moderate rotational velocities, the rotational states 
of the system are closely connected to the Mott transition. 
Since much of the qualitative physics 
is expected to be brought out already in two dimensions, the study is 
carried out in 2D. 

A gas of spinless bosonic atoms in an optical lattice is known to be 
well described by the Hubbard model \cite{jaksch1998}. 
In the presence of 
rotation, the Hamiltonian is \cite{scarola2007,wu2004}
\begin{align}
H - \Omega L_z -\mu N = -t\sum_{<ij>} \ad_i a_{j} e^{-i\phi_{ij}}
\nonumber\\
+\frac{U}{2} \sum_{i}\ad_i\ad_i a_i a_i +
\sum_i \left[\frac12\left(\tilde\omega^2 - \tilde\Omega^2 \right)r_i^2-\mu\right] \ad_i a_i,
\label{hamiltonian}
\end{align}
where $i,j$ label the lattice sites, $<\!ij\!>$ denotes nearest 
neighbors, and $\rr_i$ is the position vector of site $i$ with length 
$r_i$. $U$ is the on-site interaction strength, 
$t$ is the tunneling matrix element, and $\mu$ is the chemical potential. 
The rotation introduces a phase $\phi_{ij}$. 
An alternative Hamiltonian was proposed in Ref.\ \cite{bhat2006b} using 
a corotating ansatz for the single-particle orbitals; 
however, the quantitative differences would be 
exceedingly small in the present case. 
We state energies in units of the recoil energy $E_R$ and distances in 
units of the lattice constant $d=\lambda/2$ where $\lambda$ is the 
wavelength of the lattice beams, and for definiteness the Wannier 
functions are approximated by Gaussians. In this 
approximation 
$t=V_0^{3/4}\pi^{5/4}[1-(2/\pi)^2]\exp(-\pi^{3/2}\sqrt{V_0}/4)/4$, 
and $U = V_0^{3/4}a/\sqrt{2}$, where $V_0$ is the depth of the lattice 
potential and $a$ is the $s$-wave scattering 
length of the atoms. For the calculations, we have assumed $^{87}$Rb atoms 
with a scattering length $a=5.77$nm and a square lattice 
made out of two pairs of counterpropagating beams with wavelength $\lambda=780$nm. 
We define $\tilde\Omega=\pi\hbar\Omega/\sqrt{2}$, 
where $\Omega$ is the rotational frequency, and 
$\tilde\omega=\pi\hbar\omega/\sqrt{2}$, where $\omega$ is the 
frequency of the harmonic magnetic trapping potential.
Finally, the phase $\phi_{ij}=(m/\hbar)\int_{\rr_i}^{\rr_j}
d\rr\cdot({\mathbf \Omega}\times\rr)$, where in this 2D study the 
vector $\mathbf \Omega$ points out of the plane. 

The Hamiltonian in Eq.~(\ref{hamiltonian}) describes an optical lattice 
that rotates with the frequency $\Omega$.
Such a potential was realized in Ref.\ \cite{tung2006}.
Other ideas for creating a rotating state in an optical lattice 
include exciting the desired vorticity in a trapped condensate, 
and applying the optical lattice afterwards. 
Alternatively, one may stir the gas by using a time-dependent 
optical or magnetic potential \cite{madison2000}, or by phase engineering 
\cite{matthews1999,raman2001,andersen2006}, 
in the presence of a static optical lattice potential. 
However, the latter options are, rigorously speaking, not described by the 
Hamiltonian in Eq.~(\ref{hamiltonian}).

We now map out the rotational phases of the trapped system. 
The large number of parameters makes for a rich phase diagram. 
The rotation with frequency $\Omega$ gives rise to a 
centrifugal potential that competes with the confining 
potential. At the critical value $\Omega = \omega$,
the centrifugal potential balances the trap potential, and thus this 
criterion gives the upper bound for the angular velocity. 
However, the present method of calculation is not reliable in the 
fast rotating limit, so we will only consider 
the regime well away from this upper bound.
In addition, we work with a rather weak 
trapping potential, $\tilde{\omega}=0.025$, and hence the scaled 
rotational frequency $\tilde\Omega$ must be equally weak.
Invoking the analogy with a system in a magnetic field, 
$\tilde\Omega$ is the flux through each plaquette \cite{bhat2007}. 
Thus, the present study takes place in the limit of weak flux, 
precluding effects such as 
Hofstadter-butterfly physics \butterflycitation. However, 
this limit exhibits a number of interesting physical effects 
connected with the Mott transition, which will be the subject of the present paper. 
Furthermore, we deal only with moderate densities, 
keeping the chemical potential $\mu=0.5U$ for the most part and 
increasing it to maximally $\mu=2.0U$. In this way we can focus on 
vortex-lattice physics without invoking unnecessary complications such 
as alternating Mott and superfluid shells. 
Thus, at fixed $\tilde\omega$, we now explore the 
parameter space spanned by moderate ranges of $V_0$, $\tilde\Omega$, and $\mu$.

In the absence of rotation and external potential, the Hamiltonian 
(\ref{hamiltonian}) undergoes a quantum phase transition between two 
types of ground state at a critical value of the ratio of the 
tunneling and interaction parameters, $t/U$, depending 
on the chemical potential $\mu$ \cite{sachdev1999}. 
When the ratio $t/U$ is large 
enough, there is phase coherence over the entire sample which 
puts the system in its superfluid state, and a Bose-Einstein 
condensate is formed. 
For weaker tunneling, phase coherence is lost, 
number fluctuations are suppressed, 
and the number of atoms per site is locked to an integer. 
This is the Mott insulating state. 
The critical tunneling in two dimensions at unit filling was in 
Ref.\ \cite{elstner1999} calculated to be $(t/U)_c=0.060$, while 
in the mean field approximation, which we shall use here, 
one finds the lower value $(t/U)_c=0.042$.
In a trapping potential $V(r)$, superfluids and Mott insulators 
coexist in spatially separated regions, where 
the local phase is determined by the local chemical potential 
$\mu_{loc}=\mu-V(r)$ 
\cite{jaksch1998,campbell2006,bergkvist2004}. 
The size of the regions depends on the ratio $t/U$, such 
that for large enough $t/U$, the whole sample is superfluid, and 
as $t/U$ decreases, the Mott insulating regions grow.

If the atoms are set into rotation, it is the superfluid fraction that 
rotates since the Mott fraction is insulating. 
A continuous superfluid, or a trapped Bose-Einstein condensate in the 
absence of 
an optical lattice, responds to rotation by forming a lattice of 
quantized vortices with hollow cores \cite{pethick2002}. There exist 
excited states confined to the cores, but in a weakly interacting 
sample their population is modest \cite{fetter1972}. 
In a numerical study of an extended Hubbard model, with interactions 
between nearest neighbors, it was found that the superfluid forms 
vortices, and the cores are Mott insulating \cite{wu2004}. 
In other parameter regimes, interstitial vortices may 
form, with only modest suppression of the condensate density 
\cite{jaksch2003,burkov2006,goldbaum2008}. 
Yet another parameter regime was investigated in Ref.~\cite{scarola2007}. 
If almost all of the sample is Mott insulating, rotation will result 
in an edge superfluid containing only a few rotating atoms, and the 
bulk is unaffected.

\begin{figure}
\includegraphics[width=\columnwidth]{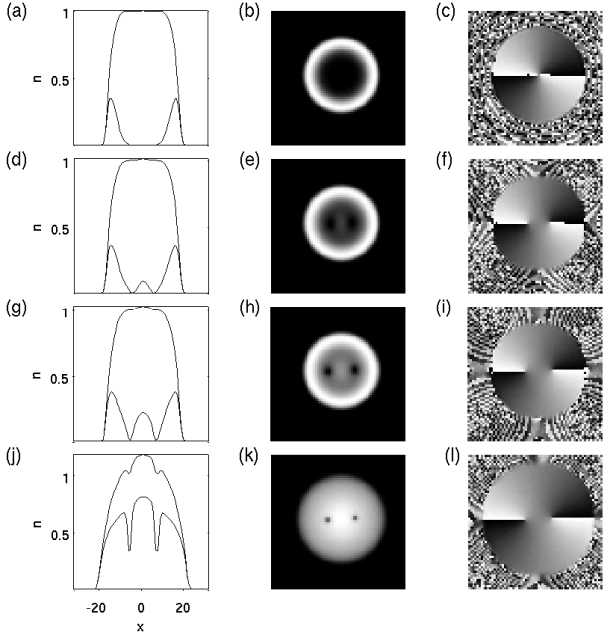}
\caption[]{Density and phase profiles 
for a system containing two phase singularities, as the Mott transition 
is crossed. 
(a), (d), (g), (j): Cross-section at $y=0$ of the dimensionless density 
$n_i$ (upper curves) and 
condensate density $n_{{\rm C}i}$ (lower curves). 
(b), (e), (h), (k): Condensate density.
Bright shades signify high density and vice versa.  
(c), (f), (i), (l): 
Phase of the condensate wavefunction $z_i$, increasing with brightness 
from 0 (black) to $2\pi$ (white).
The lattice potential height is chosen as 
$V_0=27.2$ (a-c); $V_0=26.7$ (d-f);  
$V_0=26.2$ (g-i); and $V_0=22$ (j-l). This corresponds to 
$t/U=0.0418$, 0.0447, 0.0478, and 0.0868, respectively.
The chemical potential is $\mu=0.5U$ and 
the trapping frequency is $\tilde\omega=0.025$. 
The rotating frequency 
is $\tilde\Omega=0.47\tilde\omega$ in (a-i) and $\tilde\Omega=0.63\tilde\omega$ 
in (j-l).
\label{fig:densplots_2}}
\end{figure}

Numerical calculations have been carried out on a 100$\times$100 lattice 
in the mean-field Gutzwiller 
approximation, which has proven to be reliable in a range of situations 
\cite{sheshadri1993,rokhsar1991,jaksch1998}. To solve for the rotating states, a phase winding has 
been imposed onto the nonrotating solution whereafter the energy has 
been minimized using a conjugate gradient method \cite{scarola2007}. 
5 single-particle states per lattice site were allowed for in the 
calculations. 
Solutions for the case of two phase singularities, i.e., a total phase 
winding of $4\pi$ around the circumference of the sample, are shown in 
Fig.~\ref{fig:densplots_2}. 
The Gutzwiller approximation allows for direct calculation of the 
condensate wavefunction 
$z_i=\langle a_i\rangle$ at each point; its square $n_{{\rm C}i}=|z_i|^2$ 
is denoted the 
dimensionless condensate density and is plotted alongside the dimensionless 
density, which is defined as the mean number of atoms on the site, 
$n_i=\langle \ad_i a_i\rangle$. 
The chemical potential is chosen as $\mu=0.5U$. The mean-field critical 
point for the Mott transition at $t/U=0.042$ corresponds to $V_0=27.2$.
At $V_0=26.2$, the entire system is in the 
superfluid state, and the condensate density is nonzero but contains two 
depressions which coincide with phase singularities; these are vortices. 
At these points, the total density is unity, supporting the conclusion 
of Ref.\ \cite{wu2004} that the cores are filled with Mott insulating atoms. 
As $V_0$ is increased, the condensate density in the center decreases and 
the edge-superfluid state is entered. 
At the Mott transition point $V_0=27.2$, 
the condensate density in the center is 
$n_{{\rm C}0}=3.2\cdot 10^{-6}$, and the two phase singularities are separated 
by a single lattice site. For larger $V_0$, $n_{{\rm C}0}$ 
vanishes to within numerical precision close to the center, so that the 
phase varies erratically. 
Clearly, the Mott transition is accompanied by a 
transition between the two types of rotational state. 

Since the Mott transition is second order, there is no surface 
tension associated with the phase boundary and hence no tendency for the 
vortices to stick together and form giant vortices. 
In the absence of a lattice potential, a giant 
vortex is never stable in a harmonic trap \cite{lundh2002a}, and 
apparently, the same holds for the trapped Hubbard model, 
as long as the 
whole system is in the 
superfluid phase. 

Moving further from the Mott transition, i.e., decreasing the lattice depth, 
the vortex cores are observed to decrease in size. When a vortex core is 
smaller than the lattice constant, one enters the regime of interstitial 
vortices, where the depletion of the condensate is modest. An example 
is shown in Fig.\ \ref{fig:densplots_2}(j-l). 
The reason is that this system is discrete, and the vortices naturally 
tend to form in the interstices, just as observed in most lattice 
models. In other words, on a plaquette of $2 \times 2$ 
lattice sites the phase may wind by $2\pi$, and the condensate 
density does not have to vanish on the sites because the phase 
singularity is in the void between them. A discernible vortex core 
forms when the balance between the interaction energy and the centrifugal 
energy is such that a large vortex core is favored. In a trapped 
Bose-Einstein condensate, the size of a vortex core is given by the 
healing length, $\xi=\sqrt{t/(Un_{{\rm C b}})}$, where $n_{\rm C b}$ 
is the mean 
condensate density away from the vortex \cite{pethick2002}. In the 
vicinity of the Mott transition, the formula cannot be trusted 
quantitatively, but the dependence on $n_{\rm C b}$ provides 
a qualitative explanation of why the vortex core size decreases when 
one moves away from the Mott transition.

\begin{figure}
\includegraphics[width=\columnwidth]{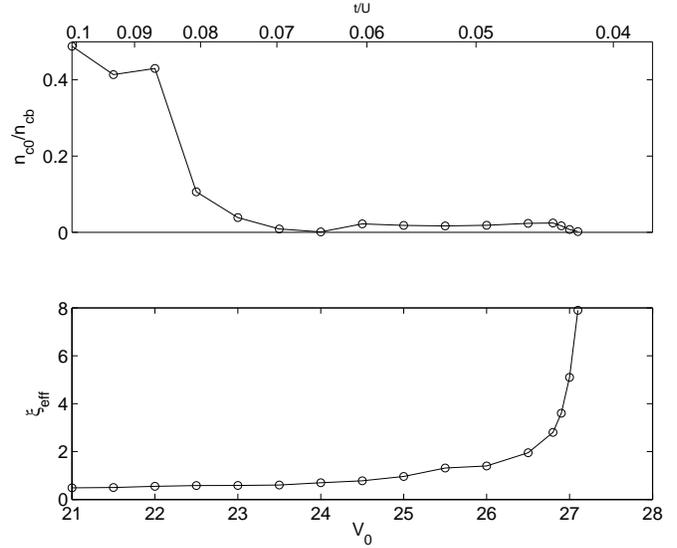}
\caption[]{(a) Size $\xi_{\rm eff}$ of a vortex core as the lattice depth 
$V_0$, and hence the tunneling over on-site energy, $t/U$, is varied. 
(b) Ratio of core to bulk condensate density, $n_{{\rm C}0}/n_{\rm C b}$, 
as a function of $V_0$ and $t/U$. 
The chemical potential is $\mu=0.5U$, the trapping frequency 
$\tilde\omega=0.025$,
and the rotational frequency is $\tilde\Omega=0.5\tilde\omega$.
\label{fig:coresize_v}}
\end{figure}

\begin{figure}
\includegraphics[width=\columnwidth]{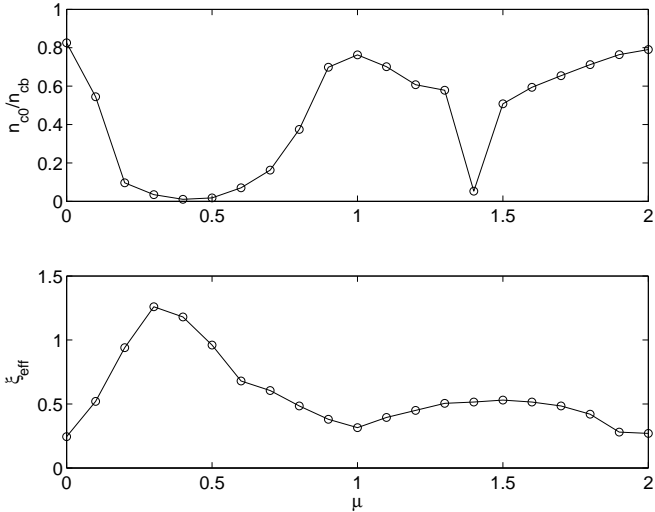}
\caption[]{(a) Size $\xi_{\rm eff}$ of a vortex core as a function of 
chemical potential $\mu$. 
(b) Ratio of core to bulk condensate density, $n_{{\rm C}0}/n_{\rm C b}$, as a 
function of $\mu$. 
The lattice depth is kept at $V_0=25E_R$, which implies $t/U=0.0564$, 
the trapping frequency $\tilde\omega=0.025$,
and the rotational frequency is $\tilde\Omega=0.5\tilde\omega$.
\label{fig:coresize_mu}}
\end{figure}

The size of the core of a single vortex has been calculated by fitting 
the central part of the condensate density profile $n_{{\rm C}i}$ to the 
the interpolation formula \cite{pethick2002,fetter1965}
\beq
n_{{\rm C}i} = n_{\rm C b}\frac{(i_x-i_{x0})^2+(i_y-i_{y0})^2}{(i_x-i_{x0})^2+(i_y-i_{y0})^2+\xi_{\rm eff}^2},
\enq
where $i=(i_x,i_y)$ is the Cartesian coordinate of a lattice site and 
$i_{x0},i_{y0}$ is the location of the center of the vortex.
Two series of calculations are made, one where the chemical potential 
is fixed at $\mu=0.5U$ and the lattice depth is varied; and one where 
$\mu$ is varied with the lattice depth fixed at $V_0=25$. 
The results for $n_{C0}$ and $\xi_{eff}$ as functions of $t/U$ is 
shown in Fig.\ \ref{fig:coresize_v}, and as functions of $\mu$ 
in Fig.\ \ref{fig:coresize_mu}. 
For simplicity, the angular frequency has been fixed at 
$\tilde\Omega=0.5\tilde\omega$, 
although the central singly quantized vortex is not the global energy 
minimum over the whole range. 

The crossover from Mott-core vortices to interstitial vortices is 
most clearly seen in 
Fig.\ \ref{fig:coresize_v}(a),
where the 
condensate density in the vortex core $n_{{\rm C}0}$ is recorded. 
In the Mott phase i.e., for $t/U < 0.042$, the condensate density in the  
center is zero to within numerical accuracy. As soon as the superfluid 
phase is entered, $n_{{\rm C}0}$ increases to a small but finite 
value. However, there is a second, reasonably sharp transition, 
where $n_{{\rm C}0}$ abruptly rises to a value of about half the 
condensate density outside the core. This transition is seen to 
occur when the fit parameter $\xi_{\rm eff}$ has dropped to a value well 
below one lattice site. The abrupt transition suggests that we can regard 
the interstitial vortices and Mott-core vortices as distinct 
rotational phases.
In Ref.\ \cite{goldbaum2008},  a homogeneous system with periodic boundary 
conditions and a large vortex density was studied.
A similar abrupt transition was found there 
between vortices centered on sites and vortices centered on 
plaquettes. 
However, as was noted in Ref.\ \cite{goldbaum2008}, since energy 
differences between different types of ground state may be small close 
to the Mott transition, it may very well be that the exact phase diagram 
is substantially different from that presented here.

In Fig.\ \ref{fig:coresize_mu}, the slightly 
oscillatory dependency of $n_{{\rm C}0}$ as a function of $\mu$ reflects the proximity in 
phase space to an alternating sequence of Mott insulating regions 
centered around $\mu=0.5, 1.5$, etc. \cite{sachdev1999,bergkvist2004}.
It is clearly seen that away from the Mott transition, the core size 
$\xi_{\rm eff}$ decreases and the condensate density in the core 
$n_{{\rm C}0}$ jumps to a value of the same order as the bulk condensate density, 
but close to the Mott insulating regions it is significantly suppressed.

\begin{figure}
\includegraphics[width=\columnwidth]{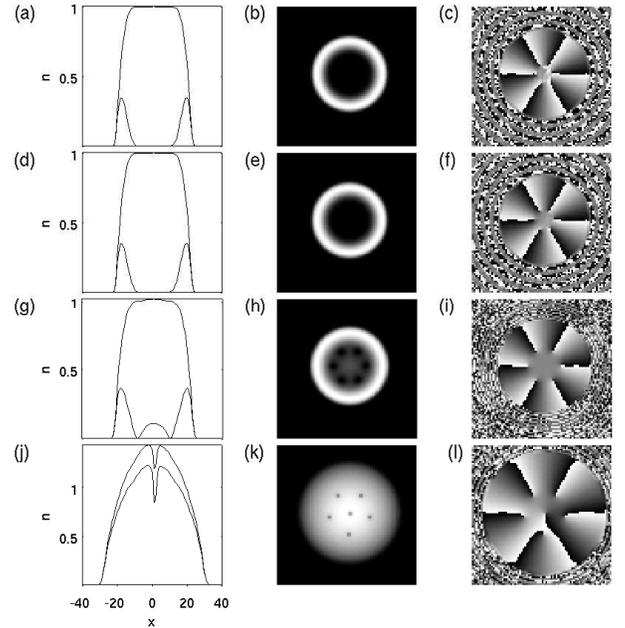}
\caption[]{Density and phase profiles for a system with six phase 
singularities. 
The quantities displayed are as in Fig.\ \ref{fig:densplots_2}.
(a-c): $V_0=27.4$, $\Omega/\omega=0.68$, $t/U=0.0407$.
(d-f): $V_0=27.2$, $\Omega/\omega=0.68$, $t/U=0.0418$.
(g-i): $V_0=26.7$, $\Omega/\omega=0.70$, $t/U=0.0447$.
(j-l): $V_0=20$, $\Omega/\omega=0.77$, $t/U=0.1176$.
The chemical potential 
is $\mu=0.5U$. 
Note that the parameters are slightly different from those in 
Fig.\ \ref{fig:densplots_2}.
\label{fig:densplots_6}}
\end{figure}

As the angular velocity is increased, the minimum-energy state of the 
system contains an increasing number of vortices. 
In Fig.\ \ref{fig:densplots_6} minimum-energy configurations with six 
phase singularities are plotted far from and close to the Mott 
transition. 
In the superfluid phase, a rudimentary vortex lattice is formed, 
resembling those found in trapped condensates 
(as well as superconductors and liquid helium, of course). 
The vortices are filled with Mott insulating atoms so that the 
total density profile is not strongly affected. 
It is seen that the lattice changes from fivefold to sixfold symmetry 
for $V_0>22$; this is further discussed below. 
At the Mott transition, $V_0=27.2$, 
the phase singularities do not seem to have molten together -- the shortest 
distance is about 5 lattice sites -- but the central density 
is $n_{{\rm C}0}=8\cdot 10^{-6}$, and as $V_0$ is further increased, 
the phase begins to vary erratically in the center, as seen in 
Fig.\ \ref{fig:densplots_6}(c). 
The figure supports the conclusion that the Mott transition marks the 
transition between a vortex-lattice state and an edge superfluid. The 
rotating force does not seem to 
shift the critical point 
at least for the rather modest 
values of $\Omega$ considered here. 

\begin{figure}
\includegraphics[width=\columnwidth]{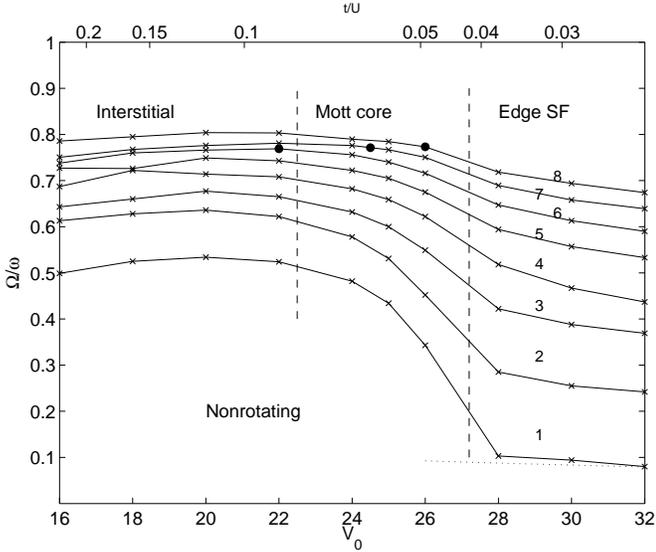}
\caption[]{
Critical frequencies $\Omega_{C,q}$ for rotational 
states in the 
Hubbard model, as a function of the optical lattice potential 
depth $V_0$. The chemical potential and external trapping 
potential are fixed at $\mu=0.5U$ and $\tilde{\omega}=0.025$. 
The numbers denote the total circulation of the state that is 
thermodynamically favorable in the indicated interval.
The dotted line is the prediction of Eq.\ (\ref{critfreqmott}).
The horizontal dashed lines mark the transitions between the 
rotational phases (see text). 
The dots mark the transitions between vortex lattices of different 
symmetry, for a total of 6, 7, and 8 quanta, respectively, in order 
of increasing $V_0$. 
\label{fig:phasediagram}}
\end{figure}

The critical frequencies $\Omega_{C,q}$ for the thermodynamic stability of 
a state containing $q$ phase singularities is calculated by comparing 
the energies of different stationary solutions of the Hamiltonian 
(\ref{hamiltonian}), computed by choosing different phase windings 
as the initial condition for the energy minimization. 
The result for the 
lowest values of $q$ is shown in Fig.\ \ref{fig:phasediagram}. 
It is known from trapped Bose-Einstein condensates that the critical 
frequency does not directly decide whether a rotating state can be 
created, since that creation depends on stability properties of 
surface excitations \cite{recati2001}. For a 
complete picture, the phase diagram 
shown in Fig.\ \ref{fig:phasediagram} should be supplemented with 
an analysis of the dynamical process for exciting vortices, but 
such an analysis is not attempted here.
 
Some features of the phase diagram can be understood quantitatively.
Deep in the Mott phase, rotation results in an edge current. 
Since the superfluid forms an annulus around the central Mott 
insulating core, the energy cost for creating a circulating current 
-- a phase winding -- is small and the critical angular velocity for 
exciting such a current will be accordingly small. 
In order to estimate the critical angular velocity, assume that the 
rotation does not appreciably change the number of superfluid 
atoms, but only induces a phase winding. The superfluid atoms, whose 
number we denote $N_C$, reside in a thin shell of some thickness $d$ 
whose radius is $R$, the size of the system. This radius is obtained 
from the condition that the external potential matches the chemical 
potential: in dimensionful units, 
$R^2=2\mu/[m(\omega^2-\Omega^2)]$. 
The velocity at radius $R$ is $v=(\hbar/m)q/R$, where $q$ is the 
phase winding. The energy associated with the 
rotation is just equal to half the squared velocity 
times the number of superfluid atoms, $E_{\rm rot}=N_Cmv^2/2$, while 
the total angular momentum is $L=N_CmvR$. Hence the angular 
velocity for stabilization of $q$ units of angular momentum is given by 
the solution to the equation 
\beq
\label{critfreqmott}
\tilde\Omega_{C,q} = \frac{\pi\hbar}{\sqrt{2}E_R}\frac{E_{\rm rot}}{L} = 
\frac{1}{\sqrt{2}\pi}
q\frac{\tilde\omega^2-\tilde\Omega_{C,q}^2}{2\mu},
\enq
where we have returned to dimensionless units. 
It is seen in Fig.~\ref{fig:phasediagram} that the estimate holds very 
well for the first critical frequency. However, it should be noted that the 
very close quantitative agreement is probably fortuitous, since the 
higher critical frequencies do not agree to the same level of precision.

For a Bose-Einstein condensate, it is known that all the critical frequencies 
should approach the trap frequency as the 
ratio of kinetic energy to interaction energy is increased \cite{pethick2002}. 
The reason why the critical frequency does not increase as $V_0$ is 
decreased below approximately $V_0=20$ is that the system enters the 
interstitial vortex phase, 
where estimates for Bose-Einstein condensates cannot be trusted. Besides, 
this study is performed at a fixed ratio $\mu/U$, 
and the total number of bosons increases as a function 
of $V_0$ at fixed $\tilde\Omega$. The transition between the 
Mott-core and interstitial vortex phases at $V_0=22.5$ is indicated 
in Fig.\ \ref{fig:phasediagram}.

The solid circles in the phase diagram mark the transitions from ring 
configurations of vortices, for large $V_0$, to lattice-like vortex 
arrays for smaller $V_0$, as exemplified in Fig.\ \ref{fig:densplots_6}.
This transition 
is most likely due to competition between two energies. The tendency for 
the vortices to arrange themselves on a ring arises from the 
``accidental'' fact that at the chosen 
chemical potential $\mu=0.5$, the condensate density $n_C$ develops a 
local minimum at a finite radius, which attracts the vortices. Further from 
the Mott transition the minimum weakens, and in combination with the 
repulsion between vortices it leads to the more familiar vortex lattices. 
By adding rotation to the already complicated phase structure of trapped 
bosons in optical lattices, 
a wealth of such different vortex configurations can be expected. 

\begin{figure}
\includegraphics[width=\columnwidth]{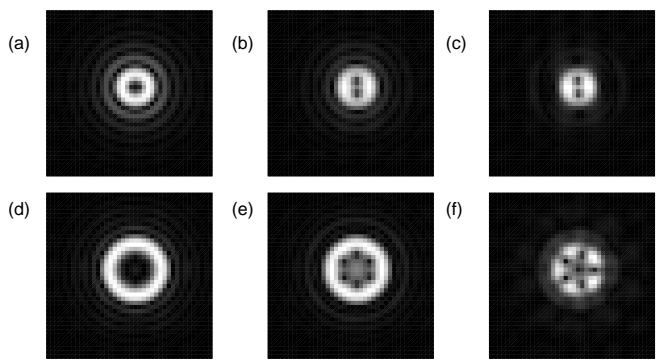}
\caption[]{Examples of density profiles after expansion of a rotating 
cloud. (a) is the expected time-of-flight image of the state 
corresponding to Fig.\ \ref{fig:densplots_2}(a-c);
(b) corresponds to Fig.\ \ref{fig:densplots_2}(g-i); 
(c) corresponds to Fig.\ \ref{fig:densplots_2}(j-l); 
(d) corresponds to Fig.\ \ref{fig:densplots_6}(d-f); 
(e) corresponds to Fig.\ \ref{fig:densplots_6}(g-i); 
and 
(f) corresponds to Fig.\ \ref{fig:densplots_6}(j-l).
\label{fig:tof}}
\end{figure}

The detection of these intriguing rotating states was discussed in 
Ref.\ \cite{scarola2007}, where it was observed that time-of-flight 
imaging can yield quantitative information about the state. 
The image of the atom cloud following a release from 
the trap will be distinctive: the Mott insulating atoms expand 
incoherently, but the Bose-Einstein condensed fraction assumes a shape 
close to the Fourier transform of the condensate wave function $z_i$; 
it is seen that the Fourier transform contains plenty of information on 
the original spatial distribution. 
Because of the periodic potential, copies of this image appear 
translated by reciprocal lattice vectors \cite{greiner2002}. 
Six examples of expected time-of-flight images are shown in 
Fig.\ \ref{fig:tof}.
In order to enhance contrast, the time-of-flight technique may be 
combined with selective removal of Mott 
insulating atoms \cite{campbell2006}. 
Time-of-flight imaging can thus easily discriminate between vortex 
lattices and edge superfluids. 
It will also directly reveal the symmetry of the vortex lattice. 
The transition between interstitial and Mott-filled vortices cannot, 
however, be easily seen in this way. in order to spot this transition, 
the method of direct detection of Mott insulating sites 
\cite{campbell2006,folling2006} 
should be used instead.

In conclusion, the rotating states for trapped bosons in an optical lattice 
have been mapped out. Three basic types of state have been identified, 
namely edge current, Mott-filled vortices and interstitial vortices. 
The transition between the two former phases coincides with the Mott 
transition. The transitions between the three types of state are found 
to be quite abrupt. 
Critical frequencies are calculated for low-lying rotating states. 
It is likely that both time-of-flight experiments and occupation-number 
sensitive detection are needed in order to detect the states discussed 
here; examples of predicted experimental signatures are calculated. 
Seeing that symmetries of vortex arrays are sensitive to the radial dependence 
of the condensate density, it is conceivable to think of vortex patterns as 
an experimental probe for the phase diagram of the trapped bosonic 
Hubbard model.

This research was conducted using the resources of High Performance 
Computing Center North (HPC2N), 
and financially supported by the Swedish Research Council, 
Vetenskapsr{\aa}det. 





\begin{thebibliography}{10}
\expandafter\ifx\csname url\endcsname\relax\def\url#1{\texttt{#1}}\fi

\bibitem{jaksch1998}
\Name{Jaksch D., Bruder C., Cirac J.~I., Gardiner C.~W. \and Zoller P.}
  \REVIEW{Phys. Rev. Lett. }{81}{1998}{3108}.

\bibitem{greiner2002}
\Name{Greiner M., Mandel O., Esslinger T., H\"a{}nsch T. \and Bloch I.}
  \REVIEW{Nature }{415}{2002}{39}.

\bibitem{campbell2006}
\Name{Campbell G.~K., Mun J., Boyd M., Medley P., Leanhardt A.~E., Marcassa L.,
  Pritchard D.~E. \and Ketterle W.} \REVIEW{Science }{313}{2006}{649}.

\bibitem{folling2006}
\Name{F\"olling S., Widera A., M\"uller T., Gerbier F. \and Bloch I.}
  \REVIEW{Phys. Rev. Lett. }{97}{2006}{060403}.

\bibitem{reijnders2004}
\Name{Reijnders J.~W. \and Duine R.~A.} \REVIEW{Phys. Rev. Lett.
  }{93}{2004}{060401}.

\bibitem{tung2006}
\Name{Tung S., Schweikhard V. \and Cornell E.~A.} \REVIEW{Phys. Rev. Lett.
  }{97}{2006}{240402}.

\bibitem{bhat2006b}
\Name{Bhat R., Peden B.~M., Seaman B.~T., Kramer M., Carr L.~D. \and Holland
  M.~J.} \REVIEW{Phys. Rev. A }{74}{2006}{063606}.

\bibitem{wu2004}
\Name{Wu C., dong Chen H., piang Hu J. \and Zhang S.-C.} \REVIEW{Phys. Rev. A
  }{69}{2004}{043609}.

\bibitem{goldbaum2008}
\Name{Goldbaum D.~S. \and Mueller E.~J.} \REVIEW{Phys. Rev. A}{77}{2008}{033629}.

\bibitem{scarola2007}
\Name{Scarola V.~W. \and Sarma S.~D.} \REVIEW{Phys. Rev. Lett.
  }{98}{2007}{210403}.

\bibitem{burkov2006}
\Name{Burkov A.~A. \and Demler E.} \REVIEW{Phys. Rev. Lett.
  }{96}{2006}{180406}.

\bibitem{niemeyer1999}
\Name{Niemeyer M., Freericks J.~K. \and Monien H.} \REVIEW{Phys. Rev. B
  }{60}{1999}{2357}.

\bibitem{jaksch2003}
\Name{Jaksch D. \and Zoller P.} \REVIEW{New J. Phys }{5}{2003}{56}.

\bibitem{oktel2007}
\Name{Oktel M.~O., Nita M. \and Tanatar B.} \REVIEW{Phys. Rev. B
 }{75}{2007}{045133}.

\bibitem{madison2000}
\Name{Madison K.~W., Chevy F., Wohlleben W. \and Dalibard J.} \REVIEW{Phys.
  Rev. Lett. }{84}{2000}{806}.

\bibitem{matthews1999}
\Name{Matthews M.~R., Anderson B.~P., Haljan P.~C., Hall D.~S., Wieman C.~E.
  \and Cornell E.~A.} \REVIEW{Phys. Rev. Lett. }{83}{1999}{2498}.

\bibitem{raman2001}
\Name{Raman C., Abo-Shaeer J.~R., Vogels J.~M., Xu K. \and Ketterle W.}
  \REVIEW{Phys. Rev. Lett. }{87}{2001}{210402}.

\bibitem{andersen2006}
\Name{Andersen M.~F., Ryu C., Clade P., Natarajan V., Vaziri A., Helmerson K.
  \and Phillips W.~D.} \REVIEW{Phys. Rev. Lett. }{97}{2006}{170406}.

\bibitem{bhat2007}
\Name{Bhat R., Peden B.~M., Seaman B.~T., Kramer M., Carr L.~D. \and Holland
  M.~J.} \REVIEW{Phys. Rev. A 
  }{74}{2006}{063606}.

\bibitem{sachdev1999}
\Name{Sachdev S.} \Book{Quantum Phase Transitions} (Cambridge University Press,
  Cambridge) 1999.

\bibitem{elstner1999}
\Name{Elstner N. \and Monien H.} \REVIEW{Phys. Rev. B }{59}{1999}{12184}.

\bibitem{bergkvist2004}
\Name{Bergkvist S., Henelius P. \and Rosengren A.} \REVIEW{Phys. Rev. A
  }{70}{2004}{053601}.

\bibitem{pethick2002}
\Name{Pethick C. \and Smith H.} \Book{Bose-Einstein Condensation in Dilute
  Gases} (Cambridge University Press, Cambridge) 2002.

\bibitem{fetter1972}
\Name{{Fetter} A.~L.} \REVIEW{Annals of Physics }{70}{1972}{67}.

\bibitem{sheshadri1993}
\Name{Sheshadri K., Krishnamurthy H.~R., Pandit R. \and Ramakrishnan T.~V.}
  \REVIEW{Europhysics Letters }{22}{1993}{257}.

\bibitem{rokhsar1991}
\Name{Rokhsar D.~S. \and Kotliar B.~G.} \REVIEW{Phys. Rev. B
  }{44}{1991}{10328}.

\bibitem{lundh2002a}
\Name{Lundh E.} \REVIEW{Phys. Rev. A }{65}{2002}{043604}.

\bibitem{fetter1965}
\Name{Fetter A.~L.} \REVIEW{Phys. Rev. }{138}{1965}{A429}.

\bibitem{recati2001}
\Name{Recati A., Zambelli F. \and Stringari S.} \REVIEW{Phys. Rev. Lett.
  }{86}{2001}{377}.

\end{thebibliography}

\end{document}